\documentstyle[aps,preprint,epsf]{revtex}
\begin{document}
\def\xx{\chi(\lambda)}
\def\bb{{1 \over M}}
\def\zu{z_1(t)}
\def\zd{z_2(t)}
\def\zt{z_3(t)}
\def\zq{z_4(t)}
\def\zi{z_i(t)}
\def\zb{{\bf z(t)}}
\def\zbb{{\bf z}}
\def\et{\epsilon(t)}
\def\eps{\epsilon}
\def\zuzd{(z_1(t),z_2(t))}
\def\wi{w_i(t)}
\def\wb{{\bf w(t)}}
\def\wj{w_j(t)}
\def\wk{w_k(t)}
\def\wnj{w_{N+j}(t)}
\def\amo{\mu_e}
\def\lio{Liouville~}
\def\sho{Schr\"odinger~}
\def\shoc{Schr\"odinger}
\def\Ha{Hamiltonian~}
\def\ehre{Ehrenfest}
\def\lbar{$\overline$}
\def\cmm{cm$^{-1}$}
\def\del{\partial}
\def\fs{$fs$}
\def\aa{\`a~}
\def\oo{\`o~}
\def\ii{\`\i~}
\def\uu{\`u~}
\def\Eq#1{Eq. (\ref{#1})}
\def\Eqs#1{Eqs. (\ref{#1})}
\def\Ep#1{(\ref{#1})}
\def\bra#1{\langle#1\vert}              
\def\ket#1{\vert#1\/\rangle}            
\def\vev#1{\langle{#1}\rangle}          
\newcommand{\bea}{\begin{eqnarray}}
\newcommand{\eea}{\end{eqnarray}}
\newcommand{\be}{\begin{equation}}
\newcommand{\ee}{\end{equation}}

\title {A Simplified Approach to Optimally
Controlled  Quantum  Dynamics}

\author {Jair Botina ~and~ Herschel Rabitz }

\address{Department of Chemistry, Princeton University, Princeton, New
Jersey 08544}

\author {Naseem Rahman}
\address{Dipartimento di Scienze Chimiche dell'Universit\aa di Trieste,
Via L. Giorgieri  1, 34127 Trieste, Italy}

\maketitle

\begin{abstract}
A new formalism for the optimal control of quantum mechanical physical 
observables is presented.  This approach is based on an analogous 
classical control technique reported previously.\cite{jairhers1} 
Quantum Lagrange multiplier functions are used to preserve a chosen 
subset of the observable dynamics of interest.  As a result, a corresponding 
small set of Lagrange multipliers needs to be calculated and they are only 
a function of time. This is a considerable simplification over traditional
quantum optimal control theory.\cite{shi1} The success of the new approach
is based on taking advantage of the multiplicity of solutions to virtually any
problem of quantum control to meet a physical objective.  A family of such simplified 
formulations is introduced and numerically tested.  Results are 
presented for these algorithms and compared with previous reported work
on a model problem for selective unimolecular reaction induced by an external 
optical electric field.
\end{abstract}

\section{INTRODUCTION} 

Different approaches and paradigms for controlling molecular motion have 
been proposed.\cite{paradigms}  Various results indicate that the final 
state distribution can be controlled in many instances.  Manipulating the 
interference between two or more routes to the same (degenerate) final state 
has been suggested to achieve control of the final state 
distribution.\cite{shap1,shap2,shap3,shap4} Another approach employs
the laser  field to guide the wave packet motion utilizing two electronic
potentials.\cite{tannor1,tannor2,kosloff} In addition to these
particular schemes, a general optimal design formalism for the
quantum control problem has been developed.\cite{wood,hrabitz,warren}
The capability of designing laser pulse shapes with this optimal 
design formalism has been demonstrated 
for many applications including selective 
excitations,\cite{shi2,judson} selective bond breaking 
for triatomic molecules,\cite{shi1} 
control of curve-crossing reactions\cite{peter1,chan} and control
of the electric susceptibility of a molecular gas.\cite{shen}
Furthermore, it  is known that multiple 
control solutions will likely exist for any particular
system.\cite{demiralp2}     
The multiplicity of solutions gives a range of flexibility for field
design which is 
especially important for adaptive feedback laboratory
control.\cite{warren,adp1}

The existence of these multiple optimal solutions provides the freedom 
to develop simplified methods to find a control field. This paper 
builds on this observation to present a different approach to quantum optimal
control theory.  The approach only requires that the feedback Lagrange
variables be  
scalars to preserve desired observable expectation values.  In
contrast, the  previous 
generic method\cite{shi1,kosloff} required the propagation of a
typically  non-linear 
\sho type equation for the Lagrange functions causing 
considerable numerical complexity.   The simplicity of the new
approach  presented here 
should allow for the study of molecular control of larger dynamical
systems.   A 
family of simplified approaches will be introduced.  The new
approaches will  be 
applied to  the selective optical dissociation of a model triatomic
molecule.   

The paper is organized as follows. In section II 
we present the general quantum control dynamics
equations and the proposed simplifications. In section III we apply the
theory to a selective unimolecular reaction.  The computational method
to solve the control dynamics equations is presented in section IV.
In section V numerical results for these methods are presented and
compared to previous results.  We discuss and
summarize the results in section VI. 

\section{THE QUANTUM CONTROL PROBLEM AND ITS SIMPLIFICATION }

In this section we present the quantum control dynamics equations for 
a molecule where the control is a laser field. The optimal control 
theory seeks a field pulse to steer the molecular motion from 
the original state to achieve a desired 
final objective. Consider the \Ha H to have the form

\be
H=H_{mol}({\bf z}) + H_{int}({\bf R},\et)
\label{Hq} 
\ee 

\noindent
where $H_{mol}({\bf z})$ is the \Ha of the undisturbed molecule,
$H_{int}({\bf R},\et)$ represents the field-molecule interaction, and 
$\et$ is the electric field vector. ${\bf z}$ is defined as an 
operator vector ${\bf z}=[z_1,z_2...z_{2N}]\equiv[{\bf R},{\bf P}] \equiv
[R_1,...,R_N,P_1,...,P_N]$ containing the coordinate ${\bf R}$ and 
congugate momentum {\bf P} operators of the system. $N$ is the 
number of degrees of freedom in the molecule.  In the dipole model, 
the interaction Hamiltonian is

\be
 H_{int}({\bf R},\et)= -\mu({\bf R})\cdot \et
\label{Hint} 
\ee

\noindent
where $\mu({\bf R})$ is the dipole moment vector.  The control design 
formalism seeks   the field $\et$.

The molecular motion evolution at time $t$ is described by 
$\ket{\psi(t)}$ which obeys

\be
i\hbar {\partial \ket{\psi(t)} \over \partial t} = H\ket{\psi (t)}
\label{schor} 
\ee

\noindent
with the initial state being $\ket{\psi (0)}$.  The knowledge of the 
wave function permits evaluation of any dynamical observables, and
in particular  $\vev{\zb}=\bra{\psi(t)}{\bf z}\ket{\psi(t)}$.  We will assume
that the control objective can be expressed in terms of $\vev{\zb}$ although 
other expectation values (e.g., bond energy, etc.) could just as well be 
treated.

\subsection{Conventional formulation of optimal control theory }

In this subsection we will briefly summarize the conventional approach 
to optimally designing fields for molecular control.  We will build on this background
and some inherent design freedom to introduce alternate and simplified 
formulations in subsection II.B. 

To design the control field $\et$ that best achieves the desired objective 
we introduce the cost functional $J[\vev{\zbb},\eps]$

\be
J[\vev{\zbb},\eps]=\Phi[\vev {{\bf z(T)}}]+ \int_0^T dt [ \ell_1(\vev{\zb})
+ \ell_2(\et) ]
\label {jsolo}
\ee

\noindent
The first part $\Phi[\vev{\bf z(T)}]$ is a 
functional that measures
the deviation from the desired physical objectives at time T.  The
second part involves the constraint  function $\ell_1(\vev{\zb)}$ that 
takes into account any restrictions on the variables $\vev{\zb}$ over
the interval $0\le t \le$ T (e.g., to avoid undesired
regions of phase space, or products, or in order to help guide the molecular
system to the desired objective.\cite{milonni})  The last part, 
with the field cost function $\ell_2(\et)$, expresses
the desire to minimize the energy fluence or possibly introduce 
other biases in the field. 

The constraint that the \sho equation be satisfied is assured through introduction
of the Lagrange multiplier vector $\ket{\lambda(t)}$ along with its complex 
conjugate.  Thus, the full cost functional is given by

\bea
\bar J[\vev{\zbb},\eps] =&& J[\vev{\zbb},\eps]
- \int_0^T dt \bra{\lambda(t)}[i\hbar {\partial \ket{\psi(t)} \over \partial t}
 - H\ket{\psi (t)}] \nonumber\\ 
&&- \int_0^T dt [i\hbar {\partial \bra{\psi(t)} \over \partial t}
 + \bra{\psi (t)}H]\ket{\lambda(t)}  
\label{jquantum}
\eea

\noindent
The minimization of $\bar J[\vev{\zbb},\eps]$ leads to a 
link between the objective functional and the optimal
solution $\et$ supplied by the Lagrange multiplier vector.

The quantum variational problem is given by $\delta \bar
J[\vev{\zbb},\eps]$=0  which minimizes the cost functional $\bar
J[\vev{\zbb},\eps]$  with respect to $\et$, $\ket{\psi(t)}$ and
$\ket{\lambda(t)}$.  The
variation with respect to $\ket{\lambda(t)}$ 
gives rise to the \sho equation \Eq{schor}.  The variation with
respect to $\ket{\psi(t)}$ leads to the following equation: 

\be
i\hbar{\partial  \ket{ \lambda(t)}\over \partial t} = H\ket{\lambda(t)} -
\sum_{i=1}^{2N} {\del l_1(\vev{\zb}) \over \del 
\vev{z_i(t)}} z_i\ket{\psi(t)} 
\label{lagr}
\ee

\noindent
This equation for the Lagrange multiplier vector has the same form as 
the Schr\"odinger equation along with an inhomogeneous term.  The final time 
condition is

\be
i\hbar\ket{\lambda(T)}= \sum_{i=1}^{2N}{\del \Phi(\vev{{\bf z(T)}}) \over \del 
\vev{{ z_i(T)}}}z_i\ket{\psi(T)}
\label{lagrT}
\ee

\noindent
The gradient of the cost functional with respect to the
field is

\be
{\delta \bar J[\vev{\zbb},\eps] \over \delta \et}= {d\ell_2(\et) \over
d\et} + 2 Im \bra{\lambda(t)} {\partial H_{int}(\bf R,\et) \over
\partial \et}\ket{\psi(t)}
\label{deltjeq}
\ee

\noindent
The solutions of the set of equations \Eqs {schor} and \Ep {lagr}-\Ep {deltjeq} 
produces the optimal field.  

In the above approach, the unconstrained cost functional $\bar
J[\vev{\zbb},\eps]$  
was introduced to conserve the \sho equation, and the control equation of motion  
for the Lagrange multiplier vector gives similar dynamics to that of 
\shoc's equation.  In this case the Lagrange multiplier vector takes into
account the state of the molecule at each instant of time in the quantum 
control process.  A key observation is that there are generally multiple 
solutions, and possibly a denumerably infinite number, to the control design 
equations.\cite{demiralp2}  We will take advantage of this flexibility below. 
The Lagrange multiplier plays the role of guiding the dynamics to a 
particular solution.

\subsection{Simplified formulation optimal control theory}

The physical cost in equation \Ep {jsolo} only depends on the expectation 
values and external interaction field and not explicitly on the wave function.  We
can take advantage of this observation to simplify the process of achieving 
control solutions.  The time dependence 
of the expectation values  is governed by the equations

\bea
{d \vev{ R_i(t)}\over dt} &&= \bra{\psi(t)}{\del H \over \del P_i
}\ket{\psi(t)} \nonumber \\
{d \vev{ P_i(t)}\over dt} &&= -\bra{\psi(t)}{\del H \over \del R_i }\ket{\psi(t)}
~~~~~~i=1,..,N. 
\label{Hamilton}
\eea

\noindent
which can collectively  be written as 

\be
{d \vev{z_i(t)}\over dt} = \bra{\psi(t)}f_i(\zbb,\et)\ket{\psi(t)}
~~~~i=1,...,2N
\label{motion}
\ee

\noindent
where the functions $f_i(\zbb,\et)$ may be readily identified as
momentum or coordinate derivatives of the Hamiltonian.  We may now 
write a new unconstrained cost functional that preserves the dynamical 
equation \Ep {motion},

\bea
\bar J[\vev{\zbb},\eps]&&= J[\vev{\zbb},\eps] \nonumber \\
- \int_0^T dt \sum_{i=1}^{2N}\lambda_i(t)\Biggl\lbrack
{d \vev{z_i(t)}\over dt} &&- \bra{\psi(t)}f_i(\zbb,\et)\ket{\psi(t)} 
\Biggl\rbrack 
\label{jquantnew} 
\eea

\noindent  
In this unconstrained functional there is an implicit dependence on 
the \sho equation, and this point will become 
important below.  It is significant to note that the functional in \Ep {jquantnew} retains
exactly the same physical objective through  $J[\vev{\zbb},\eps]$ as in 
equation \Ep {jquantum}.  Here we only alter the feedback bias towards a 
particular control solution by the choice of Lagrange functions.

The minimization of $\bar J[\vev{\zbb},\eps]$ is considered with respect to 
$\et$, $\lambda_i(t)$, and $\vev{z_i(t)}$.  The variation with respect to  
$\lambda_i(t)$ gives the quantum equation of motion for the expectation 
values, \Eq {motion}.  The variation of $\ket{\psi(t)}$ is not explicitly 
treated, but we do need to consider the interpretation of the variation 
$\delta \bra{\psi(t)}f_i(\zbb,\et)\ket{\psi(t)}$.  The various alternate optimal 
control approaches are based on different interpretations for this variation.  
We introduce the variation of $\bra{\psi(t)}f_i(\zbb,\et)\ket{\psi(t)}$, as

\bea
\delta \bra{\psi(t)}f_i(\zbb,\et)\ket{\psi(t)}&&\approx 
\sum_{j=1}^{2N}\bra{\psi(t)}{\del f_i(\zbb,\et) \over \del z_j} \delta
z_j \ket{\psi(t)} \nonumber \\
&&\approx
\sum_{j=1}^{2N}\bra{\psi(t)}{\del f_i(\zbb,\et) \over \del
z_j}\ket{\psi(t)}\vev{\delta z_j(t)} 
\label{approapp}
\eea

\noindent 
where $\vev{\delta z_j(t)}$ is 

\be
\vev{\delta z_j(t)} = \bra{\psi(t)}\delta z_j\ket{\psi(t)}
\ee

\noindent
and $\ket{\psi(t)}$ satisfies equation \Ep {schor}.  Thus, in employing the functional in
\Eq {jquantnew} an important point is that the \sho equation for $\ket{\psi(t)}$ 
is not approximated so that the true molecular dynamics is fully retained. 
Secondly, the original cost functional $J[\vev{\zbb},\eps]$  in equation \Ep {jsolo} is
retained, implying that any control solution $\et$ obtained through this alternate formulation
is just as valid as obtained the conventional way in section II.A.  
Thus, the variation in equation \Ep{approapp} should be thought of as guiding 
the design process from one valid
solution to another equally valid one.  The only question at this point is whether 
this new approach can guide the process to achieve designs $\et$ that produce 
quality control.  The computations in section III will show that the approach 
can achieve excellent results. 

Considering the above arguments we have the full variation of $\bar J[\vev{\zbb},\eps]$, as 

\begin{mathletters}
\label{alljbs}
\be
\delta \bar J[\vev{\zbb},\eps]= \sum_{i=1}^{2N}
[{\partial \Phi(\vev{\bf z(T)}) \over \partial \vev{z_i(T)}} -
\lambda_i(T)]\vev{\delta  z_i(T)}
\label{jba} 
\ee
\bea
+~\int_0^T dt\sum_{i=1}^{2N}\Biggl\lbrack{ d \lambda_i (t)\over dt} +&& 
\sum_{j=1}^{2N}\lambda_j(t) 
\bra{\psi(t)}{\partial f_j(\zbb,\et) \over \partial z_i}\ket{\psi(t)}
\nonumber \\
&&+{\partial \ell_1(\vev{\zb}) \over \partial \vev{\zi}}\Biggl\rbrack 
\vev{\delta z_i(t)} 
\label{jbb}
\eea 

\be
+~\int_0^T dt[{d \ell_2(\et) \over  d\et}
+ \sum_{i=1}^{2N}\lambda_i(t) \bra{\psi(t)}{\partial f_i(\zbb,\et) \over \partial
\et}\ket{\psi(t)}] \delta \epsilon(t).  
\label{jbc}
\ee
\end{mathletters}

\noindent
The boundary conditions at time T,

\be
\lambda_i(T) = {\partial \Phi(\vev{\bf z(T)}) \over 
\partial \vev{z_i(T)}}
~~~~i=1,...,2N,
\label{bound}
\ee

\noindent 
are obtained through the \Eq {jba}.   As, we require 
$\delta \bar J[\vev{\zbb},\eps]=0$, the second equation derived
from \Eq {jbb} is

\bea
\int_0^T dt\sum_{i=1}^{2N}\Biggl\lbrack{ d\lambda_i (t)\over dt} +&&
\sum_{j=1}^{2N}\lambda_j(t)
\bra{\psi(t)}{\partial f_j(\zbb,\et) \over \partial z_i}\ket{\psi(t)}
\nonumber \\ 
&&+{\partial \ell_1(\vev{\zb}) \over \partial \vev{\zi}}\Biggl\rbrack
\vev{\delta \zi} = 0,
\label{jbbb}
\eea 

\noindent 
and the gradient with respect to the field is

\be
{\delta \bar J[\vev{\zbb},\eps] \over \delta \et}= {d \ell_2(\et) \over d\et}
+ \sum_{i=1}^{2N}\lambda_i(t) \bra{\psi(t)}{\partial f_i(\zbb,\et) \over \partial
\et}\ket{\psi(t)}.
\label{gradfi} 
\ee

\noindent 
The Lagrange multipliers are chosen to obey the following equations

\be
{d  \lambda_i (t)\over dt} = -\sum_{j=1}^{2N}\lambda_j(t)
\bra{\psi(t)}{\partial f_j(\zbb,\et) \over 
\partial z_i}\ket{\psi(t)} -{\partial \ell_1(\vev{\zb}) 
\over \partial \vev{\zi}}~~~~~~i=1,...,2N
\label{lam1} 
\ee

\noindent
The control dynamics equations to solve the quantum control problem are
given by \Eqs {schor}, \Ep {lam1} and \Ep {gradfi} with the final conditions 
for the Lagrange multipliers given by \Eq {bound}.  The results based on this
approach will be called method I.  

The fundamental distinction between solving these equations and those
of the standard treatment in  \Eqs {schor}, \Ep {lagr}-\Ep {deltjeq} lies
in the nature of the Lagrange multipliers.  Equation \Ep{lagr} is a partial
differential equation, while \Eq {lam1} is a much simplier ordinary 
differential equation.   A key point is that the control $\et$ 
generated from the new formulation 
should be fully acceptable physically, despite the simplified form in \Eq {lam1}, 
since the \sho equation is not approximated.  The new formulation takes advantage of the 
multiplicity of solutions to steer the design process from one possible field to 
another. In some cases the control might even be similar to that found
with the full conventional control formulation (this is the case in the 
examples of Section III).  In other cases the control field  may be 
different, but importantly retention of the original cost functional 
assure a proper test of the achieved results.      

It is worthwhile to explore if further simplified formulations for the 
Lagrange multipliers can also be successfully introduced.  
One may replace  \Eq {lam1} by the following:

\be
{d  \lambda_i (t)\over dt} \approx -\sum_{j=1}^{2N}\lambda_j(t){\partial f_j(\vev{\zb},\et) \over 
\partial \vev{\zi}} -{\partial \ell_1(\vev{\zb}) \over \partial
\vev{\zi}} ~~~~~~~~~i=1,...,2N
\label{lamapp} 
\ee

\noindent
This is a classical like formulation for $\lambda_i(t)$, but quantum mechanics 
is still fully retained in solving for $\ket{\psi(t)}$ and $\vev{z_i(t)}$.  Using 
this formulation the control dynamics equations to solve are: \Eqs {schor}, 
\Ep {lamapp}, \Ep {gradfi} with  the boundary conditions \Eq {bound}.  This
approach will be called method II.

We can reformulate another cost functional based on replacing \Eq {motion}
with

\be
{d \vev{z_i(t)} \over dt} \approx f_i(\vev{{\bf z(t)}},\et)
~~~~~~~~i=1,...,2N 
\label{motionapp}
\ee

\noindent
This equation signifies that the time dependence of the expectation
values (the quantum motion behavior) resembles one classical trajectory
in the $2N$ dimensional quantum phase space.  The cost functional 
may be now rewritten as

\be
\bar J[\vev{\zbb},\eps] =
J[\vev{\zbb},\eps] - \int_0^T dt \sum_{i=1}^{2N} \lambda_i(t)
\Biggl\lbrack {d \vev{\zi} \over dt} - f_i(\vev{{\bf z(t)}},\et)\Biggl\rbrack
\label{japprox}
\ee

\noindent
The variations of $\bar J[\vev{\zbb},\eps]$ with 
respect to $\vev{z_i(t)}$, $\et$ and $\lambda(t)$ produce the control equations.  The 
variations with respect to $\lambda_i(t)$ give \Eq{motionapp}. The
variations with respect to $\vev{z_i(t)}$ gives the same boundary 
condition as \Eq {bound}.  The equations of motion for
the Lagrange multipliers are identical to \Eq {lamapp}.  However, the gradient of 
the cost functional with respect to electric field is different, 
 
\be
{\delta \bar J[\vev{\zbb},\eps] \over \delta \et}= {d \ell_2(\et) \over d\et}
+ \sum_{i=1}^{2N}\lambda_i(t) {\partial f_i(\vev{\zb},\et) \over \partial
\et}
\label{gradfi2}
\ee

\noindent
These control of equations are coupled to the \sho equation.  This approach 
will be referred to as method III below. To solve
this system of non-linear differential equations we need to know the 
expectation value for each observable in  
phase space.  The quantum control feedback equations \Ep{bound}, 
\Ep{lamapp} and \Eq {gradfi2} are identical to
classical feedback dynamics with a single trajectory, the average 
trajectory;\cite{jairhers1} a comparison between these quantum and classical 
control dynamics equations will be reported elsewhere.\cite{jairhers6} 

To summarize, the conventional approach in II.A and methods I, II and III in II.B 
should be viewed as providing alternate routes to equivalent control field designs 
consistent with the proper quantum dynamics of the molecule and the 
physical objectives.

\section{SELECTIVE CONTROL OF A UNIMOLECULAR 
REACTION }

\noindent
The conventional quantum optimal control dynamics equations \Eqs {schor}, 
\Ep {lagr}, \Ep {lagrT} 
and \Ep {deltjeq} in section II.A has been applied to a variety of problems including 
selective bond breaking through infrared
excitation.\cite{shi1}  In order to compare with the approaches described
in Section II.B, we treat the same model expressed
previously.\cite{shi1}  The selective dissociation of one bond in a linear 
triatomic molecule  is studied.  The molecule is modeled as a pair of kineticly  
coupled Morse oscillators.\cite{wilson}  The  molecular \Ha in internal 
coordinates is  

\be
H_{mol}={P_1^2 \over {2 \mu_1}}+{P_2^2 \over {2 \mu_2}} - {1 \over M_{C}}
P_1P_2 + V(R_1) + V(R_2) 
\label hmol
\ee

\noindent
where

\be
V(R_i) = D_i[1 - \exp(-\alpha_iR_i)]^2,  
\label{morse}  
\ee

\noindent
$R_1$, $R_2$ are the displacement operators of the
bonds from their equilibrium positions, $P_1, P_2$
are the conjugate momentum operators corresponding to
$R_1$ and $R_2$, the reduced masses $(amu)$ are ${1 \over \mu_1}=
{1\over 12} +  {1 \over 19}$, ${1 \over \mu_2}= {1 \over 12} +
{1 \over 35.45}$, $M_C=12amu$, and 

$$
D_1=0.1976au,~~~~~~~~ D_2=0.13705au,
\nonumber
$$
$$
\alpha_1=0.9217au~~~~~ and ~~~~~ \alpha_2=1.02725 au.
\nonumber
$$

\noindent
The dipole function of the molecule has the form

\be
\mu(R_1,R_2) = \mu_e [(R_1 + R_{o}) \exp{(-\beta R_1)}
-(R_2+R_{o}) \exp{(-\beta R_2)}]
\label{dipole}  
\ee

\noindent 
where the parameters are $\mu_e=0.3934 au$, $R_{o}=2 au$ and $\beta = 1
au$.  The  polarization of the external field $\et$ is along the axis
of the molecule.     

The model treats the selective dissociation
of bond 1 at a minimum disturbance of the non selected bond 2 along with a 
minimum fluence of the electric field.  Following the early
studies (Ref. [2]) we choose the objective function to be

\be
\Phi [\vev{\bf z(T)}]={1 \over 2} [\vev{R_1(T)} - \gamma]^2P_{f1} + {1
\over 2} \vev{P_1(T)}^2 h[-\vev{P_1(T)}]P_{f3}
\label{Phi}
\ee

\noindent
where $\gamma$ is  the ``target stretch'',
$P_{f1}$, $P_{f3}$ are positive constant weights and $h(x)$ is the  Heaviside function

\be
h(x)= \cases{
1,  &if $x \geq 0$ \cr
0,  &if $x < 0. $\cr}
\label{h(x)}
\ee

\noindent
The constraint function $\ell_1(\zb)$ is chosen as

\be
\ell_1(\zb)={1 \over 2} [\vev{R_2(t)} - \vev{R_2(0)}]^2W_2 + {1 \over
2}[\vev{P_2(t)} - \vev{P_2(0)}]^2 W_4
\label{l1}
\ee

\noindent
where $W_2, W_4$ are positive weights.  This function biases the
dynamics such that bond 2 remains minimally excited and
$\vev{R_2(t)}$ does not move far away from its initial value.  The last 
function $\ell_2(\et)$, is chosen to minimize the intensity of
the laser pulse 

\be
\ell_2(\et)={1 \over 2} \omega_e \epsilon^2(t)
\label{l2}
\ee

\noindent
where $\omega_e$ is a positive weight.

Utilizing the above criteria for the cost functional \Eq{jquantnew}, where ${\bf
z}=[z_1,z_2,z_3,z_4]$  represent the coordinate and momentum operators,  we obtain 
the corresponding set of equations \Ep {lam1} for the 
Lagrange multipliers of approach I

\begin{mathletters}
\label{alllp1s}
\be
{d \lambda_1(t)\over dt}=
\lambda_3(t)\Biggl\lbrack \bra{\psi(t)}V^{''}(z_1) \ket{\psi(t)}
- \mu_e \bra{\psi(t)}\mu''(z_1)\ket{\psi(t)} \et \Biggl\rbrack 
\label{lp1a}
\ee
\bea
{d \lambda_2(t) \over dt}=
\lambda_4(t)&&\Biggl\lbrack \bra{\psi(t)}V^{''}(z_2)\ket{\psi(t)} 
+ \mu_e \bra{\psi(t)}\mu''(z_2)\ket{\psi(t)} \et  \Biggl\rbrack
\nonumber \\ 
&&-W_2[\vev{\zd}- \vev{z_2(0)}] 
\label{lp1b}
\eea 

\be
{d \lambda_3(t) \over dt}=-{\lambda_1 (t)\over \mu_1} + \bb \lambda_2(t)
\label{lp1c}
\ee
\be
{d \lambda_4(t) \over dt}=-{\lambda_2(t) \over \mu_2} + \bb \lambda_1(t) -
W_4[\vev{\zq}-\vev{z_4(0)}]
\label{lp1d}
\ee
\end{mathletters}

\noindent
where the prime denotes the derivative with respect to $z_i$, e.g.
$V^{''}(z_i)\equiv {d^2 V(z_i) \over dz_i^2 }$.  The boundary
conditions \Eq {bound} are

\begin{mathletters}
\label{allfros}
\bea
\lambda_1(T)&&=P_{f1}[\vev{z_1(T)}-\gamma] 
\label{froa}  \\ 
\lambda_2(T)&&=0 \label{frob} \\ 
\lambda_3(T)&&=P_{f3}\vev{z_3(T)}h(-\vev{z_3(T)}) + {P_{f3} \over
2}\vev {z_3(T)}^2\delta(-\vev{z_3(T)}) \label{froc} \\ 
\lambda_4(T)&&=0  \label{frod} 
\eea
\end{mathletters}

\noindent
and the gradient of the cost functional $\bar J[\vev{\zbb},\eps]$ with respect 
of the field in \Eq {gradfi} is 

\be
{\delta \bar J[\vev{\zbb},\eps] \over  \delta \et} = \omega_e \epsilon(t) +
\mu_e [\lambda_3(t) \bra{\psi(t)}\mu'(z_1)\ket{\psi(t)} - 
\lambda_4(t) \bra{\psi(t)}\mu'(z_2)\ket{\psi(t)}]
\label{gradi}
\ee

\noindent
The minimization condition ${\delta \bar J[\vev{\zbb},\eps] \over
\delta \et} =0$ gives

\be
\epsilon_I(t) = -
{\mu_e \over \omega_e} [\lambda_3(t) \bra{\psi(t)}\mu'(z_1)\ket{\psi(t)}
 - \lambda_4(t)\bra{\psi(t)} \mu'(z_2)\ket{\psi(t)}]
\label{minie}
\ee

\noindent
The electric field depends on the two Lagrange
multipliers $\lambda_3(t)$ and $\lambda_4(t)$ and the average value of
the dipole function derivative.  Equation \Ep {minie} links the Lagrange multipliers
with the \sho equation of motion.  

Also for method II we treat the Lagrange multipliers in the same 
manner as in \Eq{lamapp}. These equations are

\begin{mathletters}
\label{alllp2s}
\be
{d \lambda_1(t) \over dt}\approx
\lambda_3(t)\Biggl\lbrack V^{''}(\vev{\zu}) 
- \mu_e \mu''(\vev{\zu}) \et \Biggl\rbrack 
\label{lp2a}
\ee
\be
{d \lambda_2(t) \over dt}\approx
\lambda_4(t)\Biggl\lbrack V^{''}(\vev{\zd}) 
+ \mu_e \mu''(\vev{\zd})\et  \Biggl\rbrack -[\vev{\zd}- \vev{z_2(0)}]W_2 
\label{lp2b}
\ee

\be
{d \lambda_3(t) \over dt}=-{\lambda_1 (t)\over \mu_1} + \bb \lambda_2(t)
\label{lp2c}
\ee
\be
{d \lambda_4(t) \over dt}=-{\lambda_2(t) \over \mu_2} + \bb \lambda_1(t) -
[\vev{\zq}-\vev{z_4(0)}]W_4
\label{lp2d}
\ee
\end{mathletters}

\noindent
The control equations \Eq{alllp2s} have the same boundary conditions
\Eq{allfros} and the same gradient of the cost functional with respect to 
electric field in \Eq{gradi}.  In this approach II the Lagrange multipliers
will differ from those of the previous approach I and therefore the
minimum solutions $\epsilon_I(t)$ and $\epsilon_{II}(t)$ are expected to be different.

The approach III has the same Lagrange multiplier equations as in the 
approach II in \Ep {alllp2s} and the gradient of the cost 
functional \Eq {japprox}  with respect to  the electric field is

\be
{\delta \bar J[\vev{\zbb},\eps] \over  \delta \et} = \omega_e \epsilon(t) +
\mu_e [\lambda_3(t) \mu'(\vev{\zu}) -
\lambda_4(t) \mu'(\vev{\zd})]
\label{gradiapp}
\ee

\noindent
The minimization condition ${\delta \bar J[\vev{\zbb},\eps] \over
\delta \et} =0$ gives

\be
\epsilon_{III}(t) = -
{\mu_e \over \omega_e} [\lambda_3(t) \mu'(\vev{\zu})
 - \lambda_4(t) \mu'(\vev{\zd})]
\label{minee}
\ee

\noindent
Note that these equations are identical to those reported
previously\cite{jairhers1} for molecular classical optimal
control evaluated over the average trajectory.  These set of quantum
control equations \Eq {alllp2s}, \Ep{gradiapp} with the boundary
conditions \Eq{allfros} can be treated consistently with the \sho
equation.  Solving the \sho equation directly, the trajectory of the
expectation values $\vev{z_i(t)}$ are obtained and are then utilized
to solve the coupled set of equations \Eq{alllp2s}, \Ep{gradiapp}.  This
is the basis for the self consistency of this approach.  

In the following section we explain the computational method used to solve the 
equations leading to $\epsilon_{I}(t)$,$\epsilon_{II}(t)$, $\epsilon_{III}(t)$.

\section{COMPUTATIONAL METHOD }

\noindent
To explain the computational methodology first consider the approach I.  The 
set of control dynamics equations to solve are  \Eqs {schor}, \Ep
{alllp1s}, \Ep{gradi} with the 
boundary conditions \Eq {allfros}.  An iterative scheme is adopted with the aim of  
finding $\epsilon_I(t)$ at a minimum cost while meeting the desired
objective.  The control 
algorithm used in this paper is the following:  

\begin{description}

\item{a)} Make an initial guess for the field $\epsilon_I(t)$.

\item{b)} Integrate \shoc's equation \Eq{schor} and calculate the 
quantum average trajectory and other expectation values necessary to 
integrate \Eq {alllp1s}.

\item{c)}Calculate the  probability of dissociation 
for each of the channels and evaluate the cost functional.  

\item{d)} Integrate \Eq {alllp1s} for the Lagrange
multipliers\cite{allen} backward in time. 

\item{e)} Calculate of the gradient of the cost functional with
respect to the electric field.

\item{f)} Update a new field from the last step. 

\noindent
The steps (b) to (f) are repeated until a converged solution is
obtained. 
\end{description}

\noindent
Similar algorithms can be constructed for the two other approaches II and
III.  In order to improve the form of $\et$ on each iteration we use the 
 conjugate gradient algorithm.\cite{numrecip}    

The \sho equation was integrated using the split operator
method.\cite{split1,split2}
This requires the propagator

\bea
U(t+\delta t,t)&&\approx \exp({-i\over 2}\int_t^{t+\delta t}[V({\bf R}) + 
H_{int}({\bf R},\epsilon(t'))]dt'/\hbar)
\exp(-i{\bf K}\delta t/\hbar)  \nonumber \\ 
&&\exp({-i\over 2}\int_t^{t+\delta t}[V({\bf R}) + 
H_{int}({\bf R},\epsilon(t'))]dt'/\hbar) 
\label{split} 
\eea 

\noindent
which evolves the wave function $\delta t$ in time.  The number of grid points 
chosen for the two coordinates
($R_1,R_2$) was 256 $\times$ 128. The range of values for $R_1$ and $R_2$
was  [-1,12]$au$ and [-1,5]$au$, respectively.  The time step $\delta
t=16.1au$ was used to propagate the wave function over a time interval
of $0.1ps$, and for times larger than $0.1ps$, $\delta
t=0.5au$ was selected.

An optical potential was introduced to absorb flux in the boundary region.  The 
new Hamiltonian is

\be
H = H_{mol} + H_{int}- iV_{opt}({\bf R})
\label{Hnew}
\ee

\noindent
where the optical potential is chosen to be

\be
V_{opt}(R_i) =\cases{
V_{o_i} {R_i - a \over b -a }, &if $a\leq R_i \leq b$; \cr
0,           &otherwise. \cr}
\label{vopt}
\ee 

\noindent
The parameters for the first bond are $V_{o1}=2a.u.$, $a=11.5a.u,
b=12.0a.u$ and those for the second bond are $V_{o2}=1a.u$, $a=4.5a.u$,
$b=5.0a.u$. 

With this scheme, we can define the dissociation probability in
various ways.  The most straightforward approach is simply to compute
the probability as

\be
P_i(t)=\int_0^{t}dt' \int_{a_i}^{b_i} dR_d {\cal J}_i(R_d,t')
\label{prodiss}
\ee

\noindent
where the ${\cal J}_i(R_d,t)$ is the flux defined as

\be
{\cal J}_i(R_d,t)={\hbar\over \mu_i}Im\bra{\psi(t)}\hat n \cdot \nabla_i \ket{\psi(t)}
\label{flux} 
\ee

\noindent
The spatial integral in \Eq {prodiss} is over the
flux dividing line between two points in the surface $a_i$ and $b_i$, and $\hat n$ is
the local unit vector normal to the dividing line.  The gradient in \Eq {flux} is
with respect to $R_1$, $R_2$ and the index $i$ refers to a specific channel in the 
photodissociation process. We can get four different channels in the 
process of dissociation, 

\be
ABC \to \cases{
A  + BC~~~~~~~dissociate~bond~1\cr
AB + C~~~~~~~dissociate~bond~2\cr
A + B + C~~~~~full~break~up\cr
ABC~~~~~~~~~~no~dissociation }
\label{reaction}
\ee

\noindent
where each process has a distint probability of occurrence.  In order 
to calculate the probability for each channel, we choose the following
values:  $P_1\equiv A+BC$ for the flux through the line [6,-1] to [6,4]; 
$P_2\equiv AB+C$ for [-1,4] to (6,4); $P_3\equiv A+B+C$ for [7,4] to [6,5] 
(all of these values are in a.u.).  The probability for $P_4\equiv ABC$ was 
calculated as

\be
P_4(t) = 1 - \sum_{i=1}^3P_i(t).
\label{P4t}
\ee

\section{NUMERICAL RESULTS  } 

\noindent
In Fig.1 we show the optimal laser pulse and its
corresponding power spectrum, utilizing the standard quantum control
equations  \Eqs{schor}, and  \Ep{lagr} to \Ep{deltjeq} of section II.A. The optimal 
laser pulse is identical to that in Ref.[2]. The power 
spectrum of the field
has only one dominant peak at $1334cm^{-1}$, which is higher than the
two fundamental frequencies $\omega_1^o=1098 cm^{-1}$ and
$\omega_2^o=923cm^{-1}$ of both Morse oscillators.  The second peak 
at $667cm^{-1}$  plays a less important role.
  
In Fig. 2 we show the optimal pulses obtained from the three different
algorithms I, II and III proposed where the guessed input field was zero.  
The optimal pulse I is produced  from the solution of the control dynamics 
equations \Ep {schor}, \Ep {alllp1s} and \Ep {gradi} 
with the boundary  condition \Eq{allfros}.  The optimal solution II is obtained with 
the Lagrange multiplier given by \Eq{alllp2s}. The result 
III in Fig. 2 is based on the control dynamics equations calculated
over one quantum trajectory utilizing \Eqs {schor}, \Ep{alllp2s},
\Ep{gradiapp} and \Eq{allfros}.  In all three cases the optimal solutions for the
electric field selectively dissociate bond 1. All three fields $\epsilon_I$, $\epsilon_{II}$
and  $\epsilon_{III}$ are strikingly similar and also closely like that in Fig. 1. The
power spectra of  $\epsilon_I$, $\epsilon_{II}$ and  $\epsilon_{III}$ are also 
similar to that in Fig. 1, except that the small low frequency peak near $700cm^{-1}$ 
is missing.

Figure 3 shows the temporal evolution of the  quantum
expectation values for the approximate bond energy, bond lengths, 
momenta and total molecular energy in the presence of the optimal field of 
Fig. 1. The analogous results for the field $\epsilon_{I}$ of Fig. 2 
is shown in Fig. 4 (the results of applying $\epsilon_{II}$ and 
 $\epsilon_{III}$ are almost the same as $\epsilon_{I}$).  The desired dissociation 
event ABC$\to$ A + BC is clearly evident.  The approximate energy for each bond
(this energy is approximate because the kinetic
coupling between the two bonds is not taken into account) shows
high  excitation in bond 1 (the dissociation energy is 0.197$a.u$)
much in contrast to bond 2 (the dissociation energy is 0.137$a.u$).
The  time evolution of the expectation values for the positions and
the momenta for the two bonds  also indicate selective dissociation
of bond 1.  The expectation values of the interaction energy and
the total energy of the molecule as a function of time  are shown in 
Fig. 5 for the conventional field in Fig. 1 and the approach I (the results
of II and III are essentially the same as the latter case). 
We observe that the interaction energy has its maximum value around 
the time $t=0.055ps$ in all the cases.

Table 1 shows the probability for the four reaction channels using 
the four optimal pulses of Figs. 1 and 2.  More than of 50\% of the
dissociation occurs for the desired channel ABC $\to$ A+BC with 
less than 1\% for the channel ABC $\to$ C+AB in all cases.  The low values
for the dissociation probabilities for the reactions ABC $\to$ C+AB
and ABC $\to$ A+B+C demonstrates the high degree of control achieved
for the new approaches to designing controls over quantum motion.  The number 
of iterations to achieve the same level of convergence with the various methods 
is essentially the same as shown in table 2.  However, the computational 
saving of methods I, II or III over that of conventional 
quantum control is nearly a factor of two as the overhead is very small for solving the
Lagrange multiplier equations with the new  methods.  The potential for further 
computational saving will be discussed in the conclusion section.

\section{CONCLUSIONS  }  

\noindent
In this work we have presented a new general formalism for the control of 
quantum observables. Illustrations were carried out to compare a 
family of related methods. The conventional reference approach standardly 
utilizes a full Lagrange
multiplier state vector introduced to preserve the \sho equation.  The 
Lagrange multipliers in the new approach are scalars introduced to 
preserve the average dynamics.  A family of related approaches I, II, III  
was presented, where each member corresponded to a distinct treatment of 
the Lagrange multiplier equations.

The four algorithms presented in this work were applied to a triatomic
molecule model studied previously.  The numerical results demonstrate
the selective dissociation of the stronger bond 1, while bond 2 remains
only weakly excited.  The approximate bond energy and the temporal
evolution of the bond length and momentum expectation values demonstrate 
selective dissociation for each of the optimal fields.  We also compared 
the expectation values of the total molecular energy with the interaction 
energy as a measure of the efficiency of the control process.
From these observations we deduce that the three new algorithms I, II, III 
proposed here give excellent and very similar results.  The dissociation probability 
of the desired bond 1 was more than 50\%, while for bond 2 the dissociation was 
less than 1\%. 

With these excellent results in evidence, a central question is why such a 
serious alteration of the Lagrange multiplier equations is successful.  Three 
factors are operative here: 

\begin{description}

\item{(1)} The proper quantum dynamics of the molecule through $\ket{\psi(t)}$ is
fully retained.

\item{(2)} The cost functional retains the goal of achieving the original target objective.

\item{(3)} There are typically multiple (if not an infinite number) of possible 
control solutions in any physically well posed molecular control problem. 
\end{description}

\noindent 
The role of the Lagrange multipliers is to provide feedback and guide 
the dynamics to an acceptable solution.  Considering the three points above, 
the alternative formulations we introduced for $\ket{\lambda(t)}$ just serve to take 
us from one acceptable solution to another.  The striking similarity of the reference 
pulse and those of methods I, II and III also suggests that the field minimization 
space of the cost functional is not locally distorted to a significant degree.  
The ultimate saving with the 
new methods resides in their simplification of the traditional strongly
coupled two point boundary value problem for $\ket{\psi(t)}$ and 
$\ket{\lambda(t)}$.  The alternate methods for $\ket{\lambda(t)}$ greatly reduce 
the complexity of this task.  The different dynamics 
for $\lambda(t)$ in methods I, II and III give almost the same fields (actually, fields II
and III are numerically the same) which in turn are very similar to that of the 
conventional approach.  This strong similarity may not always occur in other problems,
but its presence here clearly indicates the wide latitude in treating the feedback 
process in control field design.  Capitalizing on this flexibility, by the approaches
suggested here or other related ones, could greatly simplify the molecular control
design process.

In the present paper the new Lagrange variable acted to preserve
the dynamics for the molecular bond length and momentum expectation values.
The same logic could also be applied to preserve additional or distinctly 
different observable expectation values besides $\vev{z_i(t)}$. Analogous 
problems of this type 
also arise in more traditional engineering applications of control
theory, and it would be interesting to apply these reduced Lagrange
multiplier concepts in such cases.

The ultimate significance of the findings in this paper is suggested by considering the
work in the context of the analogous classical study\cite{jairhers1} and recent tracking 
control studies.\cite{tracking1,tracking2,tracking3,tracking4}  
Tracking is relevant here, as it operationally replaces 
the feedback role of the Lagrange multiplier by an expression for the control field explicitly in 
terms of the system wave function.  In essence, the present paper introduces what may appear to be 
serious approximations for the feedback Lagrange multipliers; but, in fact, 
the resultant control is well-achieved, and in some cases, the field is strikingly similar to 
that obtained by the full traditional feedback approach.  Similarly, the operations of 
tracking would appear to create a drastic modification of traditional feedback, yet tracking 
encompasses traditional optimal control solutions, as well as others.  All
of this work points to the observation that there is considerable freedom for introducing approximations
and direct physical guidance into the feedback aspects of quantum control.   
This feedback can be compactly expressed as

\be
i\hbar {\partial \ket{\psi(t)} \over \partial t}=
\Biggl\lbrack H_{mol} + \mu\eps(t,\vev{O},\ket{\psi(t)})\Biggl\rbrack \ket{\psi(t)}
\label{compact}
\ee

\noindent
where the field is shown to possibly depend on time explicitly, the expectation value $\vev{O}$ of 
target observable operator $O$ and the state $\ket{\psi(t)}$ of the system.  Tracking control has this form 
of feedback, and a formal solution to the Lagrange multiplier equations also leads to a similar form of 
feedback.   In the simplest case, \Eq {compact} only needs to be integrated once to achieve a design. 
The ultimate savings from these overall simplified approaches could be a factor of two, or up to 
many times that magnitude, when iteration is eliminated as in tracking and other direct feedback
approaches
(i.e. for tracking the factor of savings is $\sim 2N$ where $N$ is the 
nominal number of optimal control iterations and tipically $N>>10$). 
The insight gained from the present body of work suggests that focusing on the
physical and numerical content of the feedback should be a very fruitful direction in molecular
control theory.

This new approach may be conveniently combined with methods that 
compute the potential surface as the dynamics
proceeds.\cite{abn1,abn2,abn3,abn4,abn5} 
The Lagrange multiplier logic has 
now been applied to classical
mechanics\cite{jairhers1,jairhers2,jairhers3,jairhers4} and quantum 
mechanics in the present paper.  The same control concepts may also be
applied for semiclassical wave packet
propagation\cite{semi1,semi2,semi3} and mixed quantum/classical molecular
dynamics.\cite{kai,gerber,billing,augustin}

\acknowledgments

J. B. and H. R. acknowledge support for this work from the 
Army Research Office.

\newpage

\def\prl{{{Phys. Rev. Lett.}\ }}
\def\cpl{{ {Chem. Phys. Lett.}\ }}
\def\cp{{ {Chem. Phys.}\ }}
\def\cpc{{ {Comp. Phys. Com.}\ }}
\def\pra{{ {Phys. Rev. A}\ }}
\def\prb{{{Phys. Rev. B}\ }}
\def\pr{{{Phys. Rev.}\ }}
\def\jchp{{{J. Chem. Phys.}\ }}
\def\jpc{{{J. Phys. Chem.}\ }}
\def\jms{{{J. Mol. Struct.}\ }}
\def\jchso{{{J. Chem. Soc. Faraday Trans.}\ }}
\def\jcomphy{{{J. Comput. Phys.}\ }}
\def\ssc{{{Solid State Commun.}\ }}
\def\ss{{{Surf. Sci.}\ }}
\def\sci{{{Science}\ }}
\def\eurlet{{{Europhys. Lett.}\ }}
\def\prog{{ {Prog. Theor. Phys.}\ }}
\def\phy{{ {Physica}\ }}
\def\jpa{{ {J. Phys. A}\ }}
\def\achre{{ {Acc. Chem. Res.}\ }}
\def\arevpc{{ {Ann. Rev. Phys. Chem.}\ }}
\def\adchp{{ {Adv. Chem. Phys.}\ }}
\def\revmp{{ {Rev. Mod. Phys.}\ }}
\def\phrepor{{ {Phys. Reports}\ }}
\def\bibvecch#1#2{}{}

\newpage

\begin{figure}\caption{The optimal electric field and power spectrum for 
selective dissociation using the traditional approach of \Eqs {schor}, 
\Ep{lagr}, 
\Ep {lagrT} and \Eq {deltjeq}.  
The system is in the ground state at $t=0.$ The parameters in the
control dynamics equations are $\gamma=5$, $P_{f1}=400D_1\alpha_1^2$,
$P_{f3}={200 \over \mu_1}$, $W_2=2D_2\alpha_2^2$, $W_4={ 1 \over \mu_2}$.
The pulse duration time was $0.1ps$.  These parameters also apply to the 
following figures.}
\end{figure}

\begin{figure} \caption{Optimal electric fields: 
Method I using the \Eqs {alllp1s} to \Ep{gradi}; Method II using the Lagrange  
multiplier in \Eq {alllp2s}; Method III using the \Eqs {alllp2s} and \Ep {gradiapp} 
with the same boundary conditions.  The results of methods II and III are 
numerically identical. }
\end{figure}

\begin{figure} \caption{Time dependence of the quantum expectation values of the 
approximate bond energy, bond length and momentum for each bond 
for the optimal pulse of the Fig. 1. }
\end{figure}

\begin{figure} \caption{The same as Fig. 3  for the optimal pulse I of Fig. 2.  The 
results from the pulse II and III are similar to those of pulse I. }
\end{figure}

\begin{figure} \caption{Time dependence of the quantum expectation value of
the total molecular Hamiltonian $\vev{E_T}$ and the interaction 
term $\vev{E_{int}}$. 
a) for the reference pulse  in Fig. 1; b) for the pulse I of Fig.
2.  The results from the pulse II and III are very similar to that 
of pulse I.}
\end{figure}

\newpage

\begin{table}
\caption{Convergent value of the dissociation probability $P_{i}$ 
for the four reaction channels. The four channels are: $P_{1}$ 
for ABC$\to$ A + BC; $P_{2}$ for
ABC$\to$ C + AB; $P_{3}$ for ABC$\to$ B + A + C and $P_4$ means no
dissociation. CM means the conventional method (the pulse of the Fig. 1). For 
the approaches II and III
we obtain almost the same dissociation probability as that in approach
I. }
\vskip 0.5truecm
\begin{tabular}{@{\hspace{.4in}}ll@{\hspace{.4in}}ll@{\hspace{.4in}}ll@{\hspace{.4in}}}
Method& $P_1$ & $P_2$ & $P_3$ & $P_4$ \\
\hline
CM      &   0.560   &  1.8$\times 10^{-3}$    & 1$\times 10^{-3}$ & 0.437 \\
I       &   0.628   &  7.3$\times 10^{-3}$    & 2.7$\times 10^{-3}$&0.362\\
\end{tabular}
\end{table}
\vskip 4truecm 
\begin{table}
\caption{ Number of iterations for the four different methods 
in order to find the optimal pulse field in Fig. 1 and Fig. 2.  }
\vskip 0.5truecm
\begin{tabular}{@{\hspace{.4in}}ll@{\hspace{.4in}}}
Method &   Number of Iteractions\\ 
\hline
CM      &   17  \\
I       &   18  \\
II      &   19 \\
III     &   15  \\ 
\end{tabular}
\end{table}

\newpage 

\begin{figure} 
\begin{center}
Figure 1
\end{center}
\epsfverbosetrue
\epsffile[50 180 540 700]{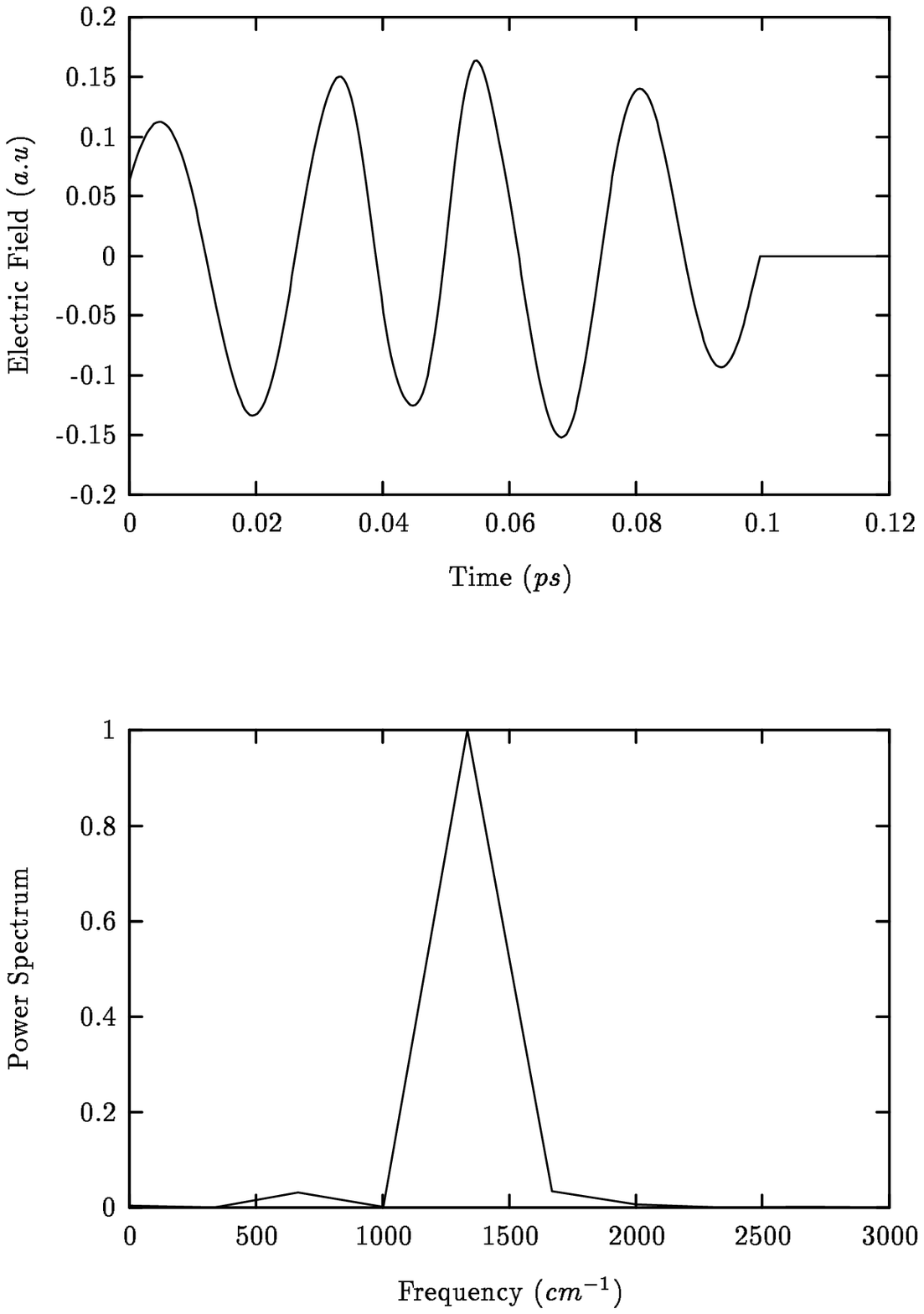}
\end{figure}

\newpage
\begin{figure} 
\begin{center}
Figure 2
\end{center}
\epsffile[50 80 540 700]{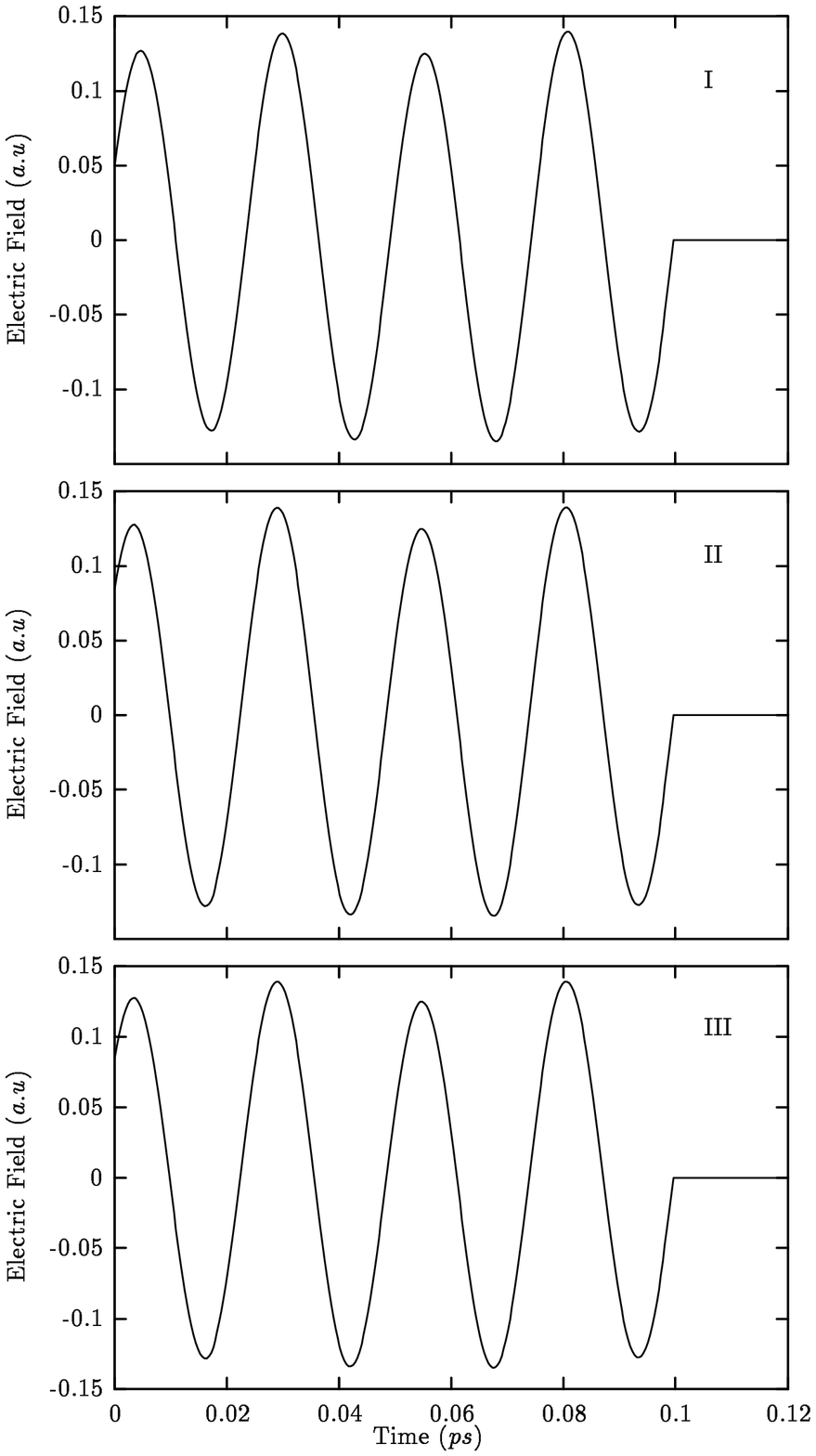}
\end{figure}

\newpage
\begin{figure} 
\begin{center}
Figure 3
\end{center}
\epsffile[50 180 540 700]{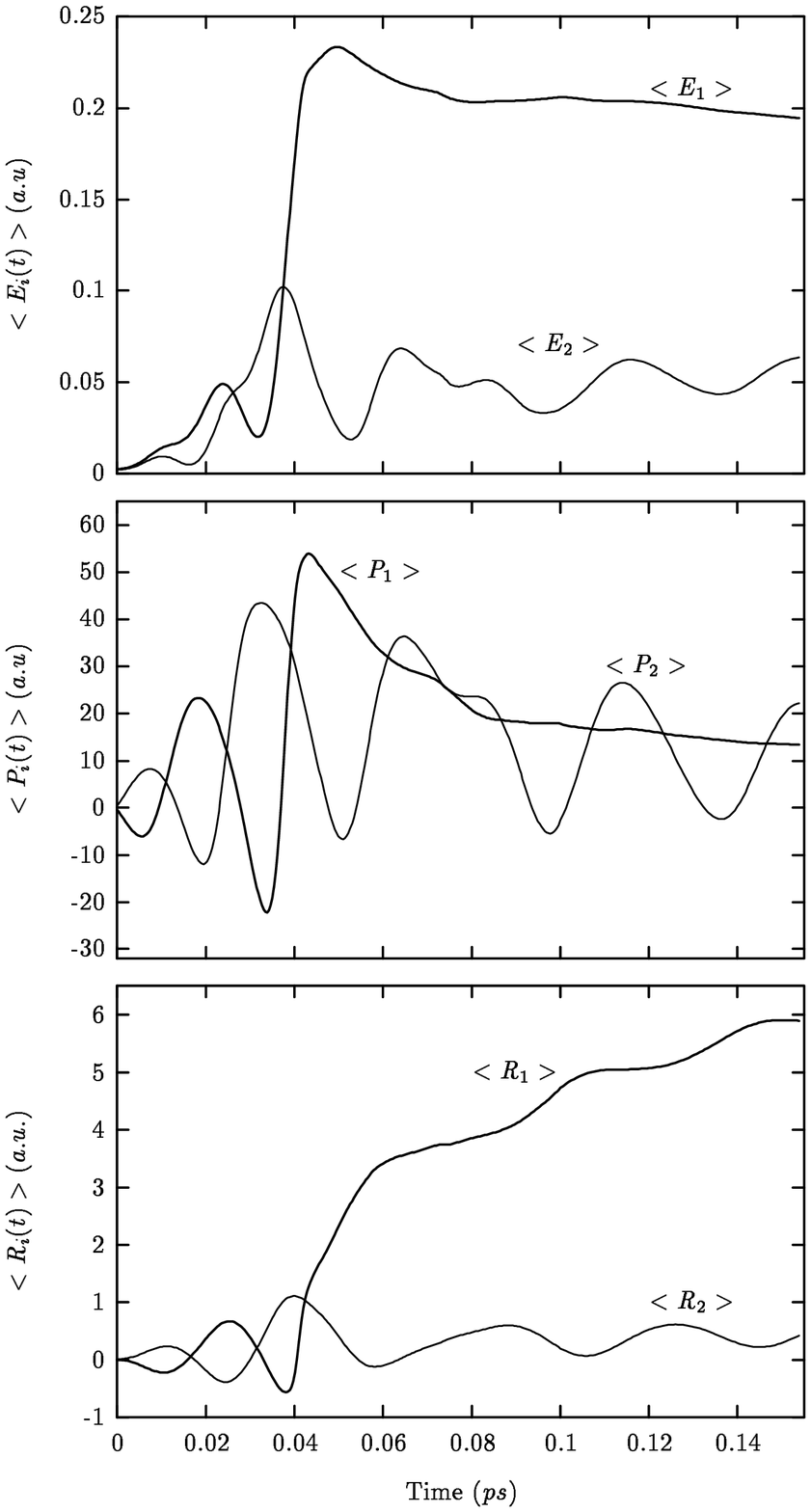}
\end{figure}

\newpage
\begin{figure} 
\begin{center}
Figure 4
\end{center}
\epsffile[50 180 540 700]{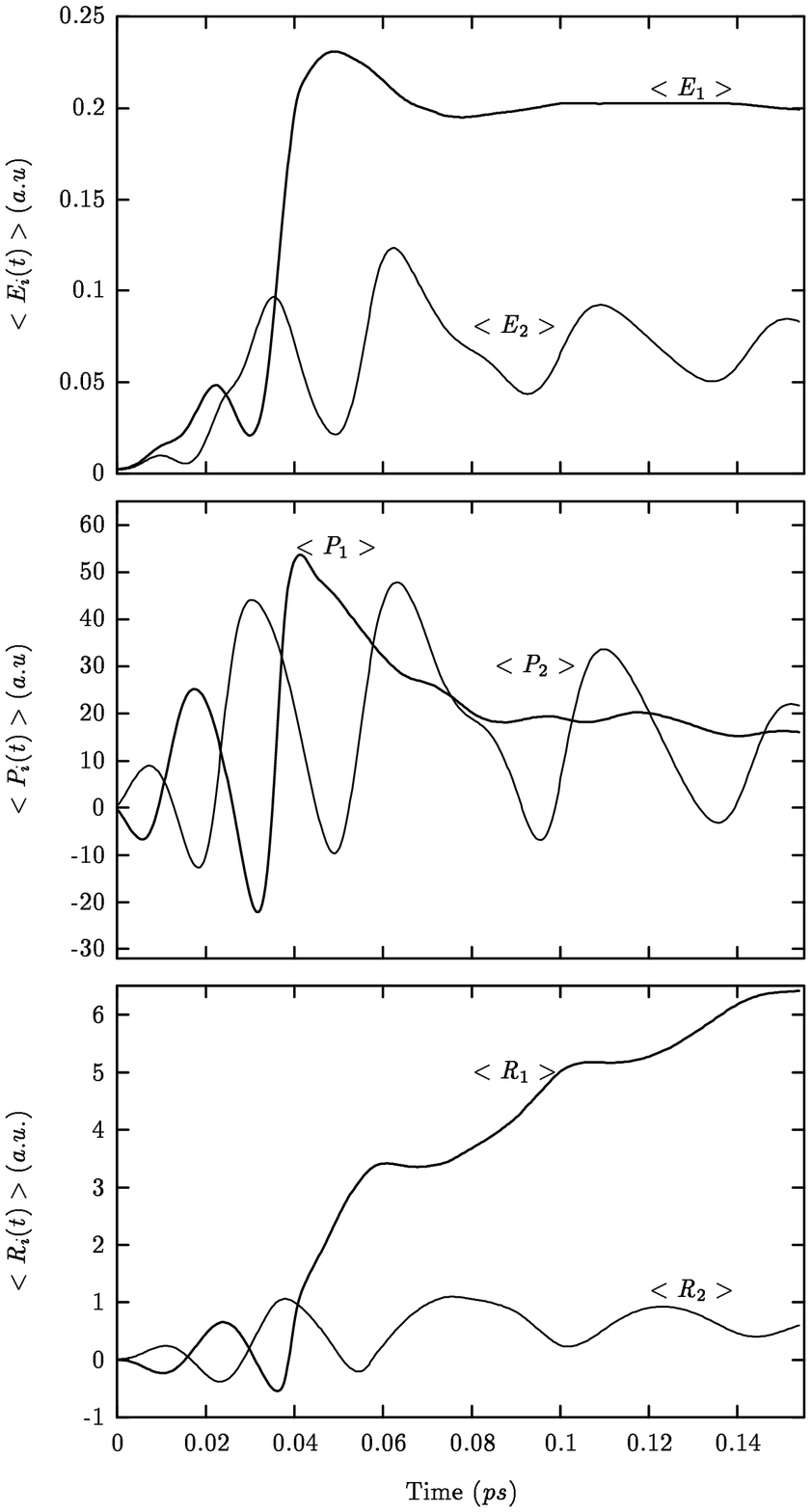}
\end{figure}

\newpage
\begin{figure} 
\begin{center}
Figure 5
\end{center}
\epsffile[50 180 540 700]{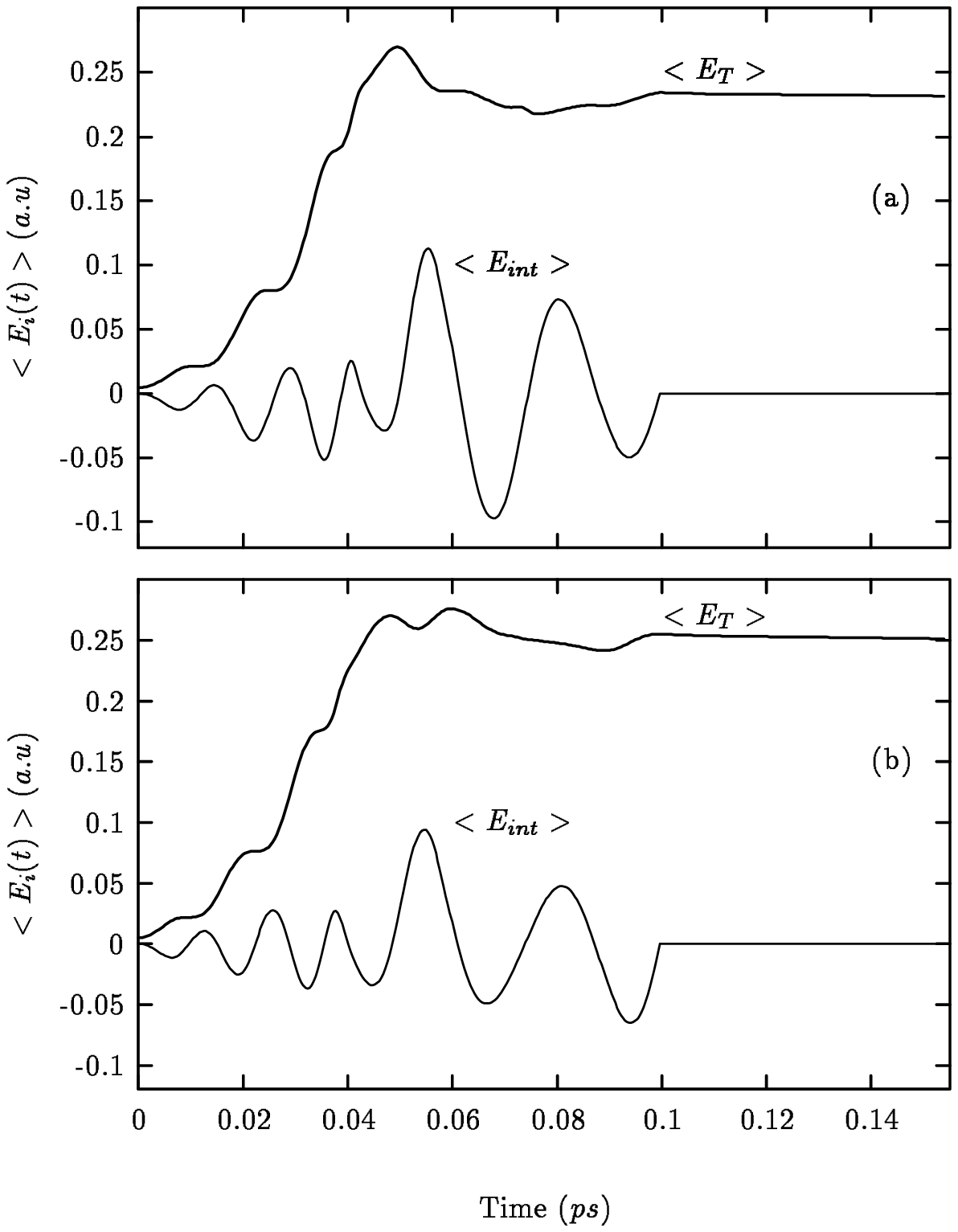}
\end{figure}

\end{document}